\documentclass[prx,aps,onecolumn,superscriptaddress,notitlepage]{revtex4-1}

\usepackage[english]{babel}
\usepackage[utf8]{inputenc}

\usepackage{color}
\usepackage[x11names,dvipsnames,svgnames]{xcolor}
\usepackage{color}
\usepackage{graphicx}
\usepackage{subfigure}
\usepackage{graphicx, subfigure}
\usepackage{fancyhdr}
\usepackage{epstopdf}
\usepackage{setspace}
\usepackage{blindtext}

\usepackage{textcomp}
\usepackage{amsmath}
\usepackage{amsfonts}
\usepackage{amssymb}
\usepackage{mathrsfs}
\usepackage{amssymb}
\usepackage{bm}  
\usepackage{txfonts}
\begin{document}

\title{Room-Temperature Continuous-Wave Frequency-Referenced Spectrometer up to 7.5 THz}

\author{Michele De Regis}
\affiliation{INO, Istituto Nazionale di Ottica-CNR, Largo E. Fermi 6, Firenze I-50125, Italy}
\affiliation{LENS, European Laboratory for Nonlinear Spectroscopy, Via N. Carrara 1, Sesto Fiorentino (FI) I-50019, Italy}
\author{Saverio Bartalini}
\affiliation{INO, Istituto Nazionale di Ottica-CNR, Largo E. Fermi 6, Firenze I-50125, Italy}
\affiliation{LENS, European Laboratory for Nonlinear Spectroscopy, Via N. Carrara 1, Sesto Fiorentino (FI) I-50019, Italy}
\author{Marco Ravaro}
\affiliation{INO, Istituto Nazionale di Ottica-CNR, Largo E. Fermi 6, Firenze I-50125, Italy}
\affiliation{LENS, European Laboratory for Nonlinear Spectroscopy, Via N. Carrara 1, Sesto Fiorentino (FI) I-50019, Italy}
\author{Davide Calonico}
\affiliation{INRIM, Istituto Nazionale di Ricerca Metrologica, Strada delle Caccie 6, Torino I-10135, Italy}
\author{Paolo De Natale}
\affiliation{INO, Istituto Nazionale di Ottica-CNR, Largo E. Fermi 6, Firenze I-50125, Italy}
\affiliation{LENS, European Laboratory for Nonlinear Spectroscopy, Via N. Carrara 1, Sesto Fiorentino (FI) I-50019, Italy}
\author{Luigi Consolino}
\affiliation{INO, Istituto Nazionale di Ottica-CNR, Largo E. Fermi 6, Firenze I-50125, Italy}
\affiliation{LENS, European Laboratory for Nonlinear Spectroscopy, Via N. Carrara 1, Sesto Fiorentino (FI) I-50019, Italy}
\email{michele.deregis@ino.it}

\date{\today}

\begin{abstract}

The lack of coherent room-temperature sources in the whole terahertz spectral window (0.3-10 THz) has significantly hampered the growth of scientific and technological applications in this range. Among them, high-precision frequency measurements of molecular transitions play a central role but remain an open challenge. Here, room-temperature generation and detection of continuous-wave, broadly tunable, narrow-linewidth THz radiation are presented, and their application to high-resolution spectroscopy in the broad 1-7.5 THz spectral range is demonstrated. This result has been achieved by implementing a Cherenkov phase-matching scheme into a channel waveguide in a nonlinear crystal. This simple approach, entirely based on robust telecom technology, unprecedently merges in a single source an ultra-broad continuous wave spectral coverage and a state-of-the-art accuracy ($\sim 10^{-9}$) in molecular transition center determination. 
\end{abstract}

\maketitle
\date{\today}

\section{Introduction} 
The dream of a metrological-grade, continuous wave (CW) source spanning the whole THz region is about 30 years old \cite{rev_luigi,lewis}, and dates back to the Tunable Far Infrared (TuFIR) difference-frequency-generation (DFG) approach \cite{evenson,zink}. This kind of source had proven able to perform high resolution spectroscopy up to 6 THz in a “standard” configuration \cite{nolt}, with two examples at frequencies as high as 7.9 \cite{odashima1} and 9.1 THz \cite{odashima2}, thus challenging the most advanced present-day CW THz sources. However, its low reliability, the low emitted power and the very bulky instrumentation, have hampered for decades its widespread use. More recently, the simplicity of its DFG scheme inspired more compact designs \cite{hindle,scheller} that, however, remained far from the former record tunability range combined to metrological performance. Ultrafast technology has very recently lead to the availability of commercial THz sources for low resolution applications \cite{krok}, although even the most advanced metrological-grade systems could only be used as frequency reference \cite{luigi_natcomm,barbieri} for THz Quantum Cascade Lasers (QCLs)-based high-accuracy spectrometers \cite{saverio_prx}. Nonetheless, in the last few decades, the THz window has fostered a number of scientific and technological applications, spanning fields as diverse as imaging in biomedicine \cite{massi}, homeland security \cite{federici}, environmental monitoring \cite{luigi_sensors}. \\
In this framework, THz molecular spectroscopy has a key role and high precision measurements on different molecular transitions have been reported in the lower part of the THz spectrum, e.g. at 0.29 THz \cite{skryl}, 0.62 THz \cite{hsieh}, 1.1 THz \cite{puzzarini}, 2.5 THz \cite{saverio_prx,bray,luigi_sat} and 3.3 THz \cite{hagelshuer}. Recent technologies, such as Frequency-Comb-based Photo-mixers \cite{hindle,mouret} and Quantum Cascade Lasers (QCLs) \cite{saverio_prx,luigi_sat,hagelshuer} have demonstrated their capability to perform metrological grade THz spectroscopic measurements.  However, each of these set-ups has its own potentials and drawbacks: for example, spiral-antenna emitters, used in photo-mixing synthesizers, show a spectrum limited to 3 THz \cite{hindle}. Similar limitations occur for QCL-based spectrometers. Despite having proven a very narrow emission linewidth \cite{miriam,ravaro}, QCLs are intrinsically limited by the semiconductor material at high frequencies, with a present-day limit set at about 4.7 THz in CW operation \cite{schrottke,ohtani}, while the widest tunability range is restricted to 330 GHz \cite{qin}. These limits, added to the need of cryogenic cooling, have hampered the range of precise spectroscopic measurements in the THz window. The most recent designs of DFG-based QCLs are trying to remove these limitations, reaching a single-mode output power as high as 4 $\mu$W, tuning in the 2-4.35 THz range \cite{lu}, and, on a different device, a spectral linewidth of 125 kHz at 20 $\mu$s \cite{luigi_io}. However, high-resolution applications have not been attempted on this kind of sources, yet.\\
In the set-up presented here, we preserve the simplicity of a DFG scheme, combining it with rugged, reliable, compact and high-power telecom lasers. Our THz source simultaneously achieves a three octaves spectral coverage (1-7.5 THz), thanks to the wide tunability range of the pump lasers, and $\mu$W-range power levels, exploiting a waveguided Cherenkov emission scheme. Finally, the presented source is applied to high-accuracy, because traceable to the primary frequency standard, molecular spectroscopy, resulting in a state-of-the-art $10^{-9}$ accuracy in the line centre determination, still preserving a room-temperature detection scheme.

\section{Waveguided Cherenkov nonlinear generation}
Most of the drawbacks of direct lasing devices, photoconductive antennas or frequency multipliers can be overcome with a nonlinear generation approach.  Thanks to their anharmonic response to optically induced excitations, a wide range of materials, i.e. semiconductors like GaAs \cite{rice,zheng}, GaSe \cite{shi} and GaP \cite{tanabe,chang}, organic materials (DAST) \cite{kawase1,kawase2} and ferroelectric crystals such as LiTaO3 \cite{jepsen} and Lithium Niobate (LN) \cite{hebling,stepanov}, can be exploited for THz generation. In fact, if we consider two electric fields, $E_1$ and $E_2$ respectively, at angular frequencies $\omega_1$ and $\omega_2=\omega_1+\omega_{THz}$, whose difference falls in the THz range, when they are mixed in such non-linear media, coherent radiation at other frequencies is produced, including the difference of the incoming ones $\omega_{THz}$. However, the efficiency of a difference frequency generation (DFG) process depends on the wavevector mismatch $\Delta {\mathbf k}  ={\mathbf k_1}-{\mathbf k_2}-{\mathbf k_{THz}}$   among the three fields, limiting the volume of the nonlinear medium effectively involved in the frequency down-conversion.\\
In particular, assuming both the pump fields $E_1$ and $E_2$ propagating in the same direction (with velocity $c/n_{pump}$ ),  only a finite thickness of the nonlinear medium is involved in the emission process,  namely the coherence length given by $L_c=2\pi/\Delta k =2\pi c/([\omega_{THz} (n_{pump}-n_{THz} cos \phi) )]$.\\
Here, we have indicated as $n_{THz}$ the THz refractive index in the nonlinear medium, and with $\phi$ the angle formed by ${\mathbf k_{THz}}$ with respect to ${\mathbf k_1}$ and ${\mathbf k_2}$.
Different techniques have often been used to restore the phase matching condition in a collinear emission geometry (i.e. with $\phi=0$) taking advantage of the material birefringence \cite{shi} or, as in the case of periodically-poled Lithium Niobate (PPLN) crystals, of a spatially periodic modulation of the nonlinear susceptibility \cite{sasaki}. \\
A disadvantage of collinear DFG schemes is the strong THz absorption, typically affecting all the nonlinear materials and motivating the set-up of non-collinear schemes. For example, M. Koch et al. \cite{scheller} demonstrated monochromatic THz surface emission from a PPLN crystal with a generation efficiency about $10^{-8}$ W$^{-1}$ and with an optical power in the milliwatt range. However, the resonant cavity used to increase the pump intensity up to 500 W unavoidably implies a reduction in the tuning range, which is limited to 200 GHz.\\
In order to overcome such limitations, a different strategy to maintain the phase matching condition stands in the Cherenkov emission scheme \cite{l'huillier, huiller:2007a, deregis:2018}. Indeed, even having a uniformly distributed nonlinear susceptibility, for a specific direction defined by the so-called Cherenkov angle
\begin{eqnarray}
\phi_c=arccos \left(\frac{n_{pump}}{n_{THz}}\right),
\end{eqnarray} 
the perfect phase-matching condition $\Delta k=0$ is exactly fulfilled and the THz radiation is coherently emitted along the whole crystal length L. Under these conditions, the generation efficiency $\eta$, increasing $\propto L^2$, is defined by the ratio $P_{THz}/(P_1 P_2)$ between the THz power and the product of the pump beam powers:
\begin{eqnarray}
\eta=\frac{L^2 d_{eff}^2 \omega_{THz}^2}{\varepsilon_0 c^3 \pi^2 n_{THz}n_{pump}^2 r_0^2}e^{-(\omega_{THz}/\omega_0)^2}
\end{eqnarray}
where $d_{eff}$ is the effective nonlinear coefficient, $\varepsilon_0$ is the dielectric constant of vacuum and $r_0$ is the waist of the pump laser beams.  In the previous expression, the cut-off frequency $\omega_0$ is determined by the refractive indexes mismatch and by the radius $r_0$, according to the expression $\omega_0=2c/r_0 (n_{THz}^2-n_{pump}^2 )^{-1/2}$. As a consequence, Cherenkov phase-matching in bulk nonlinear crystals is limited to few THz, and, in order to achieve a good efficiency in the high frequency part of the THz spectrum, the pump beam waists have to be in the few-microns range. For this reason, we propose a novel setup capable of extending the pulsed operation of K. Suizu et al. \cite{Suizu:09} to CW broadband operation, confining the pump beams radiation in a single-mode linear waveguide.\\  
As a matter of fact, to achieve Cherenkov phase-matching, it is crucial that the pump fields propagate in the nonlinear medium faster than the THz light, as in the case of infrared pumps and LN crystals. Because of this latter feature and of the strong nonlinearities \cite{roberts,harada}, combined with the high photorefractive damage threshold \cite{bryan,furuya} and the low absorption coefficient, in the order of $10^{-4}$ cm$^{-1}$ in the infrared spectral region \cite{schweig}, 5\% MgO-doped LN is widely used for THz experiments. 

\section{Results and discussion}
In our experimental setup, shown in fig.1, two commercial CW laser systems (Toptica DL Pro and MOT/DL Pro), emitting at telecom wavelengths, provide narrow linewidth (tens of kHz) radiation, tunable over several THz. A fraction of the light available from these seeding lasers is sent to a wavelength meter (Bristol model 721) that provides a measurement of the pump/signal emission frequencies with an accuracy of about 100 MHz, while their frequency separation can be continuously adjusted in the range of interest, from 1 to 8 THz. The two monochromatic sources seed two distinct fiber amplifiers: one amplifier (A1 – IPG Photonics EAR-10K-C-SF, power up to 11 W) operates between 1540 nm and 1565 nm wavelength, providing a tunability range of about 3 THz. For the other amplifier (A2) two models can be used, operating in two different narrow bands: 1573-1577 nm (IPG Photonics EAR-7K-1575-SF, with a power up to 7 W) and 1603-1607 nm (IPG Photonics EAR-3K-1605-SF, with a power up to 3 W), respectively. The two available combinations allow generation in the 0.97-4.57 THz and 4.54-8.12 THz spectral ranges, respectively, while the DFG process transfers the spectral properties of the seed telecom lasers to the THz beam. As shown in the following, the THz frequency scans, necessary to perform molecular spectroscopy, are performed in different ways, depending on how broad and how accurate the scan is needed. \\
The output beams of the two fiber amplifiers are overlapped by a non-polarizing beam splitter, while two independent optical systems (not shown in fig.1), one for each pump laser, allow for a fine adjustment of the beam shape and polarization, in order to maximize the mode matching for an efficient coupling to the channel waveguide. \\
\begin{figure*}[tbph]
\begin{center}
\includegraphics[scale=0.65, keepaspectratio]{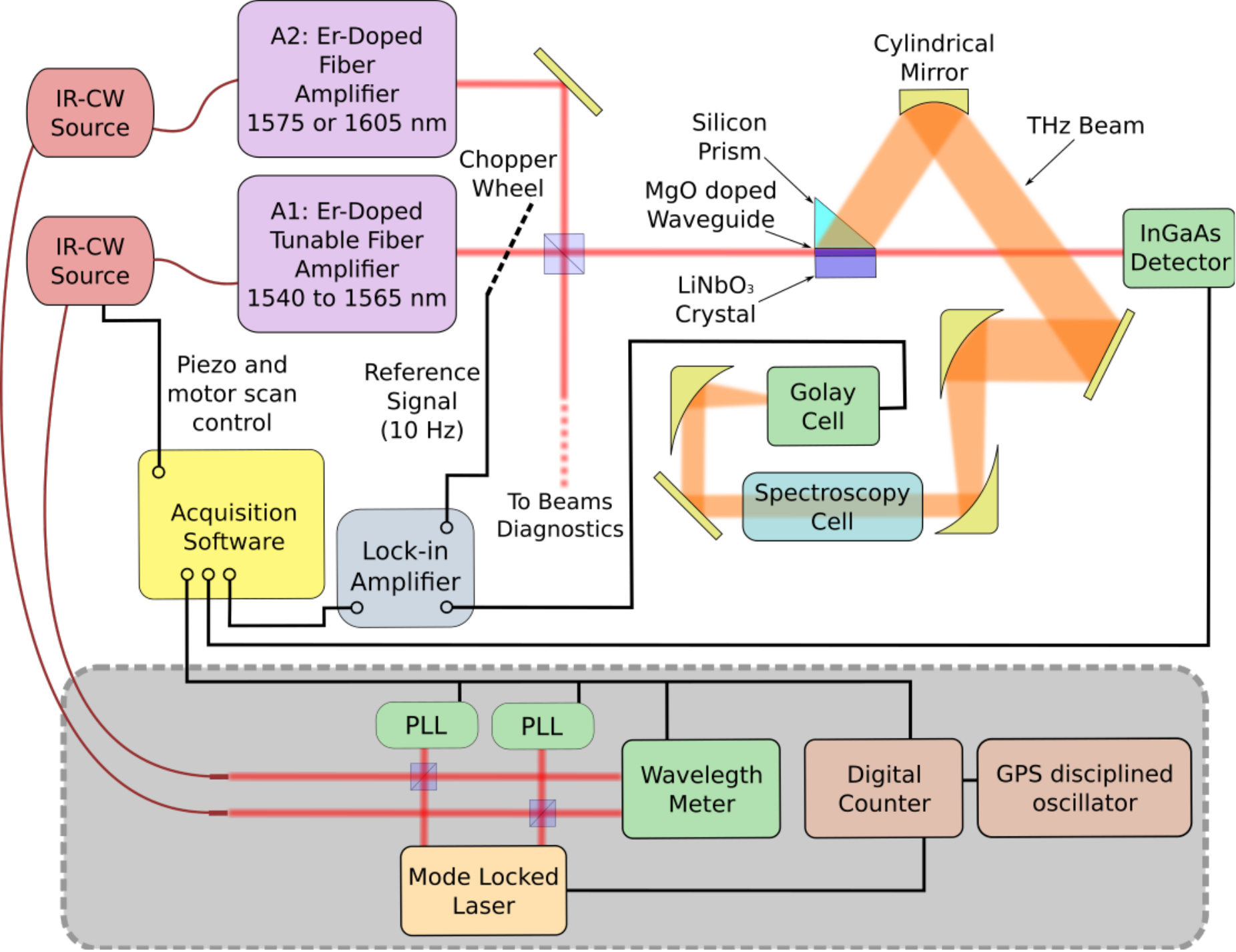}
\end{center}
\caption{Schematics of the experimental set-up. Black lines represent electric connections, brown lines represent optical fibers. The shaded grey panel represents the setup used for frequency referencing to the primary frequency standard (see section B).}
\end{figure*} 
The transverse dimensions of the mode propagating into the waveguide, i.e. the dimensions of the generating volume, are approximately 5x3 $\mu$m, so that THz absorption by the nonlinear medium is negligible. Coherent emission occurs along the whole 1 cm crystal length and, thanks to the high confinement of the guided modes, the DFG optical process is enhanced inside the waveguide. The Cherenkov angle is about 65$^o$, with small variations within the tuning range, and is not large enough to avoid total internal reflection. For this reason, a silicon prism has been positioned on top of the generating surface, ensuring (according to the Snell’s law) an efficient extraction of the THz light. A cylindrical mirror with 1'' focal length corrects the wide angular divergence of the THz beam in the direction orthogonal to the waveguide, leading to a fairly collimated THz beam, which is sent to the spectroscopic set-up.\\
The non-zero reflectivity of the crystal surfaces takes to a standing-wave modulation of the confined radiation intensity during the frequency scans, inevitably affecting the intensity of the generated THz field. Furthermore, the total pump intensity, as high as $2.4 \cdot 10^7$ W/cm$^2$, induces thermal drifts of the crystal length, whose effects are also present in the THz signal. These effects were removed from the acquisitions by monitoring the pump beams powers during the frequency scans, using a 10 Hz mechanical modulation on one of the pump beams and a fast detector (an InGaAs photodiode with 150 MHz bandwidth – Thorlabs PDA10CF-EC) placed at the waveguide output. This setup provides a continuous monitoring of the guided mode intensities and, therefore, allows retrieval of the generation efficiency $\eta$.

\subsection{SPECTRAL COVERAGE, MODE-HOP FREE TUNABILITY AND GENERATION EFFICIENCY}
The generated optical power was detected in a non-coherent scheme, using a low noise Golay-type detector (Tydex GC-1P) with a noise -equivalent power of about $10^{-10}$ W Hz$^{-1/2}$ at 10 Hz, and with a calibrated responsivity of 20 kV/W . The THz signal is modulated at 10 Hz by the mechanical chopper placed along the optical path of one of the IR pump beams, and is then demodulated by a Lock-in amplifier (Stanford Research Systems SR-830) integrating over several modulation periods. \\
In Fig.2, two different broadband scans are shown, in order to demonstrate the unprecedented CW spectral coverage achieved by this set-up. Wavelength tuning, here, is determined by the motorized rotatable grating of the external cavity of the laser seeding the A1 amplifier, Toptica MOT/DL Pro, with an accuracy better than 1 GHz. The curves reported in Figs.2a and 2b were obtained by varying the tunable laser wavelength between 1542 nm and 1563 nm with the other laser emitting at 1575 nm and 1605 nm, respectively.  \\
In order to protect the Golay detector from irreversible damage induced by residual pump laser radiation, a 1 mm thick polypropylene (PP) filter was placed before the detector itself. This ensures a 50\% transmission of the THz radiation, and a $10^{-6}$ attenuation factor for the undesired IR residual radiation. Nevertheless, it introduces broad absorptions around 3.3 THz and 6.8 THz. Both these effects are well visible in the green traces of Fig.2a and Fig.2b, respectively, showing the THz power detected in the whole generation bandwidth. A lower but flatter and broader 25\% transmission is obtained by replacing the PP filter with a 1-mm-thick germanium filter (fig. 2 blue traces).  In this condition, the generation curve presents a cut-off at 7.5 THz, that can be probably ascribed to the well-known transverse optical phonon resonance at 7.5 THz \cite{bakker}, and to the two-phonon absorption \cite{winer} arising in the few mm thick silicon prism. \\
In the first spectral band, the maximum THz power detected by the Golay cell is about 0.5 $\mu$W, while the telecom pump powers in the waveguide are 1.8 W each. This leads to a maximum generation efficiency as high as $\eta=1.5 \cdot 10^{-7}$ W$^{-1}$. A similar efficiency is found in the second band, where a 0.2 $\mu$W power is detected at around 5.2 THz, with pump powers of 1.8 W and 0.8 W, respectively. \\
\begin{figure}[tbph]
\begin{center}
\includegraphics[scale=0.32, keepaspectratio]{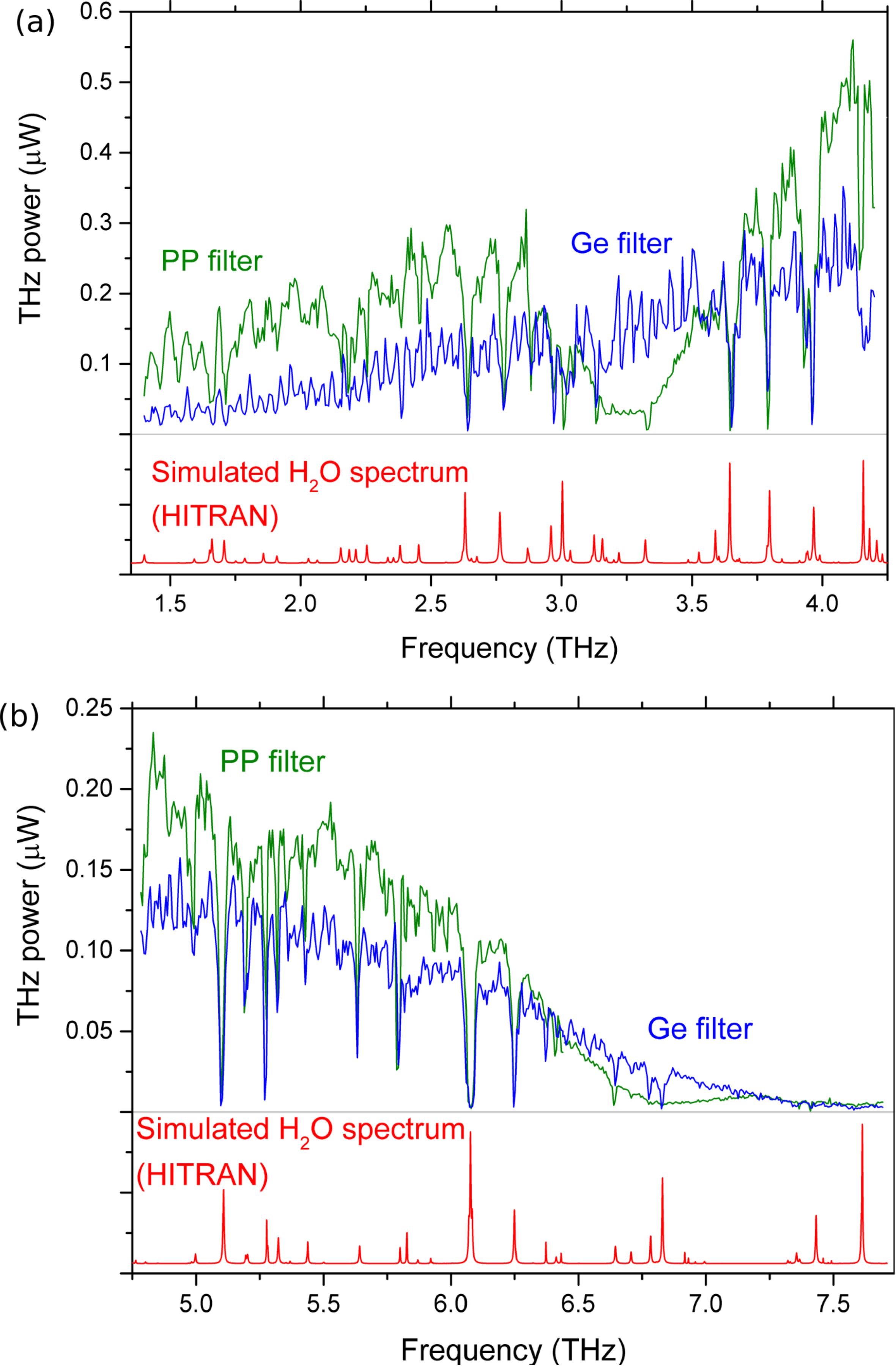}
\end{center}
\caption{Generation curves corresponding to a tunable pump laser wavelength scan from 1542 to 1563 nm. The fixed wavelength pump lasers operate at 1575 nm (plot a) )  and 1605nm (plot b)). The blue and green lines were recorded using a Ge and a polypropylene filter, respectively. The red trace shows the water absorption profiles simulated using the parameters given in the HITRAN molecular database. }
\end{figure} 
A box with dry air purging was built along the THz-wave path (not shown in Fig. 1) in order to reduce the absorption due to water vapor contained in the laboratory air. Nevertheless, water absorption lines are clearly visible in fig.2, and can be exploited as frequency calibration marks when comparing with the absorption spectrum provided by the HITRAN database (red lines in Fig. 2). The frequency step used in these acquisitions is 0.05 nm, corresponding to around 6 GHz in frequency, thus comparable with water vapor linewidths at atmospheric pressure. \\
In order to achieve a more accurate control on the emitted THz radiation, it is possible to adjust the wavelength of the laser seeding amplifier A1 using a piezoelectric tuning of the diode external cavity, driven by a temporized DC signal. To assess the performance of this tuning mechanism, the generated THz wave was collimated and sent through a spectroscopy cell, placed before the detection stage (see fig. 1), and previously filled with low-pressure (in the range 10-200 Pa) methanol vapors. \\
Several spectroscopic tests, performed along the whole generation span, demonstrate that the source provides a 40 GHz mode-hop-free operation along the whole generation bandwidth, and a tuning coefficient, necessary to calibrate the frequency scans, could be retrieved (see Appendix B for details). \\
\begin{figure}[tbph]
\begin{center}
\includegraphics[scale=0.15, keepaspectratio]{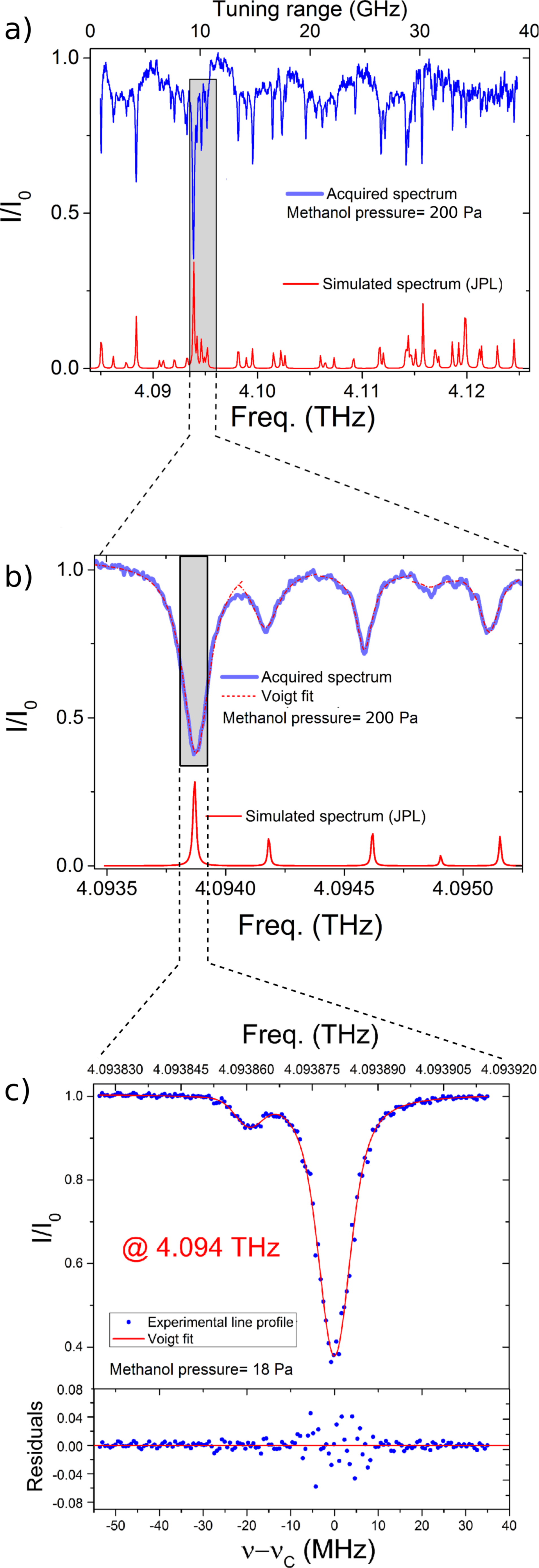}
\end{center}
\caption{Using the piezoelectric actuator on the cavity length of the CW laser seeding the A1 amplifier, different ranges of THz frequencies can be investigated: a) 40 GHz mode-hop free continuous spectrum of methanol vapors at 200 Pa, 0.1 V/step, and simulated spectrum from JPL database. b) 2 GHz scan at 200 Pa and 25 mV/step, corresponding to 6 MHz/step, Voigt fits of the recorded transitions and simulated spectrum from JPL database. c) 90 MHz zoom on the 4.094 transition of Methanol with 0.25 mV/step, corresponding to 60 kHz/step. The experimental data have been fit to a Voigt profile, while the lower panel shows the fit residuals.}
\end{figure} 
Fig. 3 shows the versatility of such spectrometer, and a comparison with a simulation based on the Jet Propulsion Laboratory (JPL) molecular database is shown. Fig. 3a shows the 40 GHz wide mode-hop-free continuous acquisition of the methanol spectrum, with a gas pressure of 200 Pa and a step of 0.1 V on the piezoelectric actuator. Fig. 3b shows a 2 GHz zoom around 4.094 THz, acquired at the same methanol pressure and with a 25 mV step, corresponding to a frequency resolution of about 6 MHz (see Appendix B for details). In order to further zoom in the most intense of these transitions, the methanol pressure was lowered down to 18 Pa, and the acquisition is shown in fig. 3c over a 90 MHz span. It was recorded with a 0.25 mV step, corresponding to about 60 kHz resolution in the THz frequency. The acquired profile was fitted to a Voigt function, which provided an uncertainty on the line center frequency $\Delta \nu_C=33$ kHz and a relative precision of $(\Delta \nu_C)/\nu_C =8.06 \cdot 10^{-9}$.  

\subsection{HIGH-ACCURACY SPECTROSCOPIC MEASUREMENTS}
In this section, the performance of our source in terms of high-accuracy spectroscopy is shown. In order to provide an absolute scale for the THz radiation, a telecom mode-locked femtosecond laser (Menlo Systems C-fiber HP) was used as a frequency ruler. The details of the experimental setup are shown in the grey panel of fig. 1.\\
The i-th tooth of the comb-like emission of the fs laser has a frequency:
\begin{eqnarray}
\nu_i=f_{off}+i\cdot f_{rep},
\end{eqnarray}
where $f_{off}$ is the carrier-envelope offset frequency, and $f_{rep}$ is the repetition rate frequency, adjustable around 100 MHz for our laser. The two CW seed lasers are mixed with the closest teeth on a fast detector (Thorlabs DET10C/M), namely the Nth tooth for the laser seeding the tunable A1 amplifier, and the Mth tooth for the laser seeding one of the two fixed amplifiers A2. The resulting beat notes are phase locked with standard loop systems (PLLs) to two synthesized local oscillators (LOs) at a frequency around 30 MHz, using the seed laser driving current.
When the two PLLs are closed, the two laser frequencies can be written as:
\begin{subequations}
\begin{equation}
\nu_M=f_{off}+M\cdot f_{rep}+f_{bM} 
\end{equation}
\begin{equation}
\nu_N=f_{off}+N\cdot f_{rep}+f_{bN},
\end{equation}
\end{subequations}
where $f_{bN}$ and $f_{bM}$ are the two LOs frequencies, and their signs depend on whether the beating tooth is at higher or lower frequency with respect to the seed. The generated THz radiation will be:
\begin{eqnarray}
\nu_{THz}=\nu_M-\nu_N=(M-N)\cdot f_{rep}+f_{bM}-f_{bN}.
\end{eqnarray}
The two LOs frequencies are generated by two synthesizers, disciplined by a GPS-Rb-OCXO (Oven Controlled Crystal Oscillator) chain frequency standard, while a disciplined frequency counter measures the repetition rate. The uncertainties on the LOs are negligible ($<10^{-11}$), while the frequency counter returns a 0.05 Hz uncertainty on the instantaneous measurement of $f_{rep}$. In this configuration, the generated THz frequency can be directly traced against the primary frequency standard. It is known with an accuracy depending on the M-N term, and is in the order of few kHz.\\
When the PLLs on the seed lasers are active, control on the generated THz frequency is achieved by tuning the mode-locked laser repetition rate. As shown in the equation providing $\nu_{THz}$, a change of $\Delta f_{rep}$ translates in a $(M-N)\Delta f_{rep}$ change in the THz frequency. Such acquisitions allow for a very fine tuning over hundreds of MHz scans, providing an absolute frequency scale for spectroscopic acquisitions. In order to test the spectrometer accuracy, a set of measurements was performed on methanol vapor transitions. Such transitions, thoroughly studied for years and tabulated in catalogues, were used as a benchmark for our set-up. \\
\begin{figure}[tbph]
\begin{center}
\includegraphics[scale=0.35, keepaspectratio]{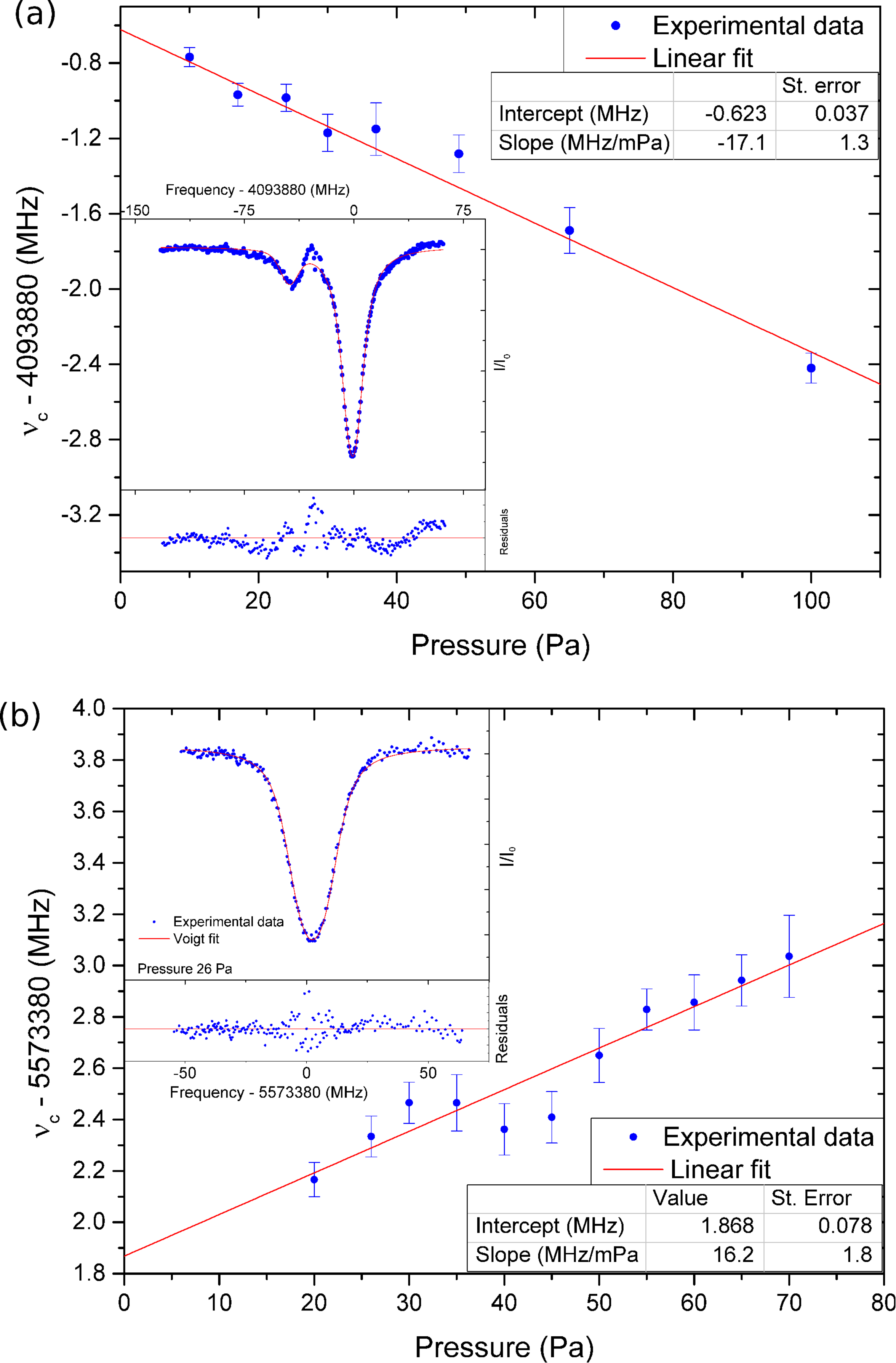}
\end{center}
\caption{Measured dependence of the centre frequency on the gas pressure for the two analysed transitions and corresponding linear fits. The intercept and slope values from the fits give the absolute frequency and the pressure shift of the considered line. The insets of the two figures show two acquisitions, with their respective Voigt fit and fit residuals, for a pressure of 26 Pa.}
\end{figure} 

The insets of figures 4a and 4b show two methanol absorption profiles, the former reporting the methanol rotational transition $(J,K,\upsilon_T )=(14,6,1\rightarrow 15,7,1)$ at 4.094 THz, the latter the transition $(J,K,\upsilon_T )=(4,4,2\rightarrow 5,5,2)$ at 5.573 THz, $\upsilon_T$ being the torsional quantum number. These two transitions lie in two distinct bands covered by the two different fiber laser amplifiers combinations. An integration time of 3 seconds was chosen as a trade-off between a good signal-to-noise ratio and an overall measurement time of about 10 minutes. In both cases, the normalized transmitted intensity $I/I_0$ is plotted as a function of frequency. Each recorded transition profile was fitted to a Voigt profile (see Supplemetary Material). In order to retrieve the absolute value of the center frequency $\nu_C$, a systematic analysis at different methanol vapor pressures was carried out, and the pressure shift was measured for both transitions: the linear fits are shown in figs. 4a and 4b. The resulting transition frequencies are 4,093,879,377(37) kHz for the former, corresponding to an absolute accuracy of $9.0\cdot 10^{-9}$, while for the latter the retrieved value of 5,573,381,868(78) kHz results in a $1.4\cdot 10^{-8}$ absolute accuracy.\\
The obtained results were then compared with previous measurements performed on the same transitions \cite{moruzzi}, with the calculations performed by the same authors, and with the JPL molecular database, based on the simulations by Xu et al. \cite{xu}. In particular, the experimental data of \cite{moruzzi} were taken at a pressure of 200 Pa, while the pressure shift was not measured. Therefore, we also added to the comparison the measured frequency values from Ref. 50 but frequency shifted according to our measured value of self-pressure shift (-17.1 MHz/mPa for the first transition and 16.2 MHz/mPa for the second).\\
Regarding the transition at 4.093 THz, the comparisons are reported in fig. 5a. The JPL database value (4,093,878,690(58) kHz) has a discrepancy of about 690 kHz with respect to our data, while the agreement between the previous measurement in \cite{moruzzi} and the value from our work significantly improves when applying the self-pressure shift to the measurements. Fig. 5b shows the comparison for the transition at higher frequency. The JPL database reports this transition with an absolute frequency more than 150 MHz away from all the other values, while our data are much closer to the measurements from \cite{moruzzi}. The self-pressure shift coefficient of the two transitions, measured thanks to the accuracy of our setup, allowed to improve the experimental data from \cite{moruzzi}, shifting the measured data closed to the theoretical ones for both transitions. As a final consideration, when comparing the measurements presented in this work and the pressure shifted measurements of \cite{moruzzi}, we retrieve for both considered transitions a discrepancy very close to 1.9 MHz. This discrepancy, if confirmed for a larger number of transitions, could be ascribed to a systematic error on one of the two experimental setups. 
\begin{figure}[tbph]
\begin{center}
\includegraphics[scale=0.35, keepaspectratio]{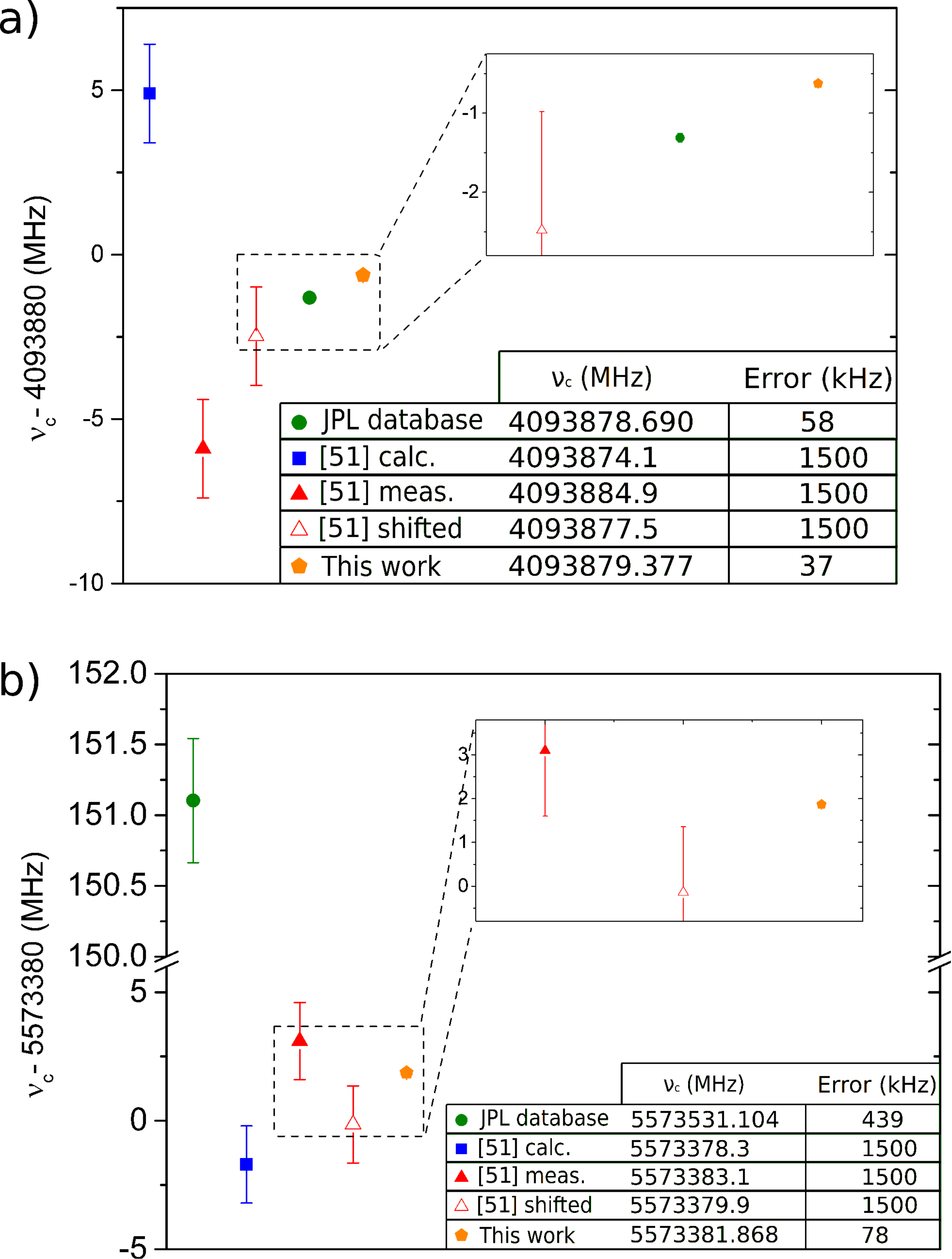}
\end{center}
\caption{The absolute frequency of the two analysed transitions can be found in the JPL molecular database (green) and both the measured (red) and the theoretically predicted (blue) values reported in ref \cite{moruzzi}. These results are compared with the measurements from the present work (orange). We also report the measured values from \cite{moruzzi}, but frequency shifted according to our measured self-pressure shift coefficient (empty red triangles)  (a) A comparison for the 4.0930 THz transition and (b) for the 5.573 THz one.}
\end{figure} 
\section{conclusions}
In conclusion, we demonstrated high-accuracy spectroscopic operation of a room temperature CW THz source based on rugged, compact and reliable telecom fiber laser technology. Our setup uniquely combines, in a single source, an ultra-broad spectral coverage, spanning from 0.97 to 7.5 THz; a $\mu$W-level emission power; an absolute frequency reference, enabling measurements of molecular transition frequencies at an accuracy $\sim 10^{-9}$. The achieved level of accuracy, if systematically reproduced over a large set of transitions of selected gases, could help to significantly improve molecular models. Indeed, significant deviations have been found for the transition frequencies measured in our work, as compared with previous results, even for a long studied molecule like methanol. Moreover, the combination of high resolution and accuracy of our set-up can help to compare previous data with present ones, by using pressure shift data. Regarding systematic effects that can affect the actual accuracy, it is worth noting that, besides the collisional shift, other effects such as Zeeman shift, ac-Stark shift, or blackbody shift are expected to contribute below the $10^{-10}$ accuracy scale, and were therefore neglected in our analysis.\\
The broad spectral coverage of the presented source is enabled by the simultaneous use of a Cherenkov emission scheme and strong light confinement in a surface nonlinear waveguide, leading to a significant broadening of the phase matching bandwidth. Moreover, thanks to the guided-wave approach, our set-up reaches CW generation efficiencies as high as $1.5 \cdot 10^{-7}$ W$^{-1}$, and grants power levels high enough for both room temperature detection and high-accuracy THz spectroscopy. Compared to the metrological performance of other state-of-the-art spectrometers, our precision is one order of magnitude better than that achieved with photo-mixing-based synthesizers \cite{hsieh}, and comparable to that of the best QCL-based systems \cite{saverio_prx}, needing however liquid-He cryogenic temperatures to operate in narrow spectral ranges. \\
The current limit to the accuracy achievable with our setup is represented by the Doppler-limited transitions. This limit can be overcome either by directly performing sub-Doppler measurements in room-temperature samples in a cell \cite{puzzarini,luigi_sat,wienold}, that could require cryogenic detection due to the power levels available, or by interrogating a cold gas sample in a molecular beam \cite{insero,santamaria}. Similar approaches, exploiting the narrower Doppler-free transitions, could get to a further improvement in the measurement accuracy of at least one order of magnitude, i.e. better than $\sim 10^{-10}$, thus taking THz measurements even closer to those that can be performed in other, well developed, spectral regions.

\section*{Acknowledgments}
Authors would like to acknowledge support from: European Union FET-Open grant 665158 `'Ultrashort Pulse Generation from Terahertz Quantum Cascade Lasers''-ULTRAQCL Project; European ESFRI Roadmap `'Extreme Light Infrastructure''-ELI Project; European Commission-H2020 Laserlab-Europe, EC grant agreement number: 654148.
\appendix
\section{Crystal plate design} 
The Y-cut 5\%-MgO-doped LiNbO$_3$ (LN) crystal plate (by HC-Photonics corp.) has  8.3x10x0.5 mm$^3$  size. Several channels (two slabs and nine channel waveguides) are patterned on top of the -Y face with a depth of $\sim$2 $\mu$m. The single-mode waveguides used in this work are oriented along the 10 mm long side and were designed for optimal light confinement of near-IR radiation. The effective mode size is about 5 $\mu$m diameter. In order to take full advantage of the nonlinear properties of LN, the orientation of the optical field polarization and the crystal optical axis were chosen parallel both to the crystal plate surface and to the input facet. 
\section{Mode-hop-free scans and tuning coefficient determination}
A number of spectroscopic tests, performed on methanol vapors, were performed along the whole generation span. These tests are useful to retrieve the spectrometer performance in terms of mode-hop-free operation, and to retrieve the frequency/voltage coefficient. Examples of such wide-span scans, with a methanol vapor pressure of 200 Pa, are shown in Figs. 3a and 3b, and are reported in the insets of fig. 6, for convenience.\\
The first and the last absorption dips, visible in the blue trace of Fig. 6a inset, respectively correspond to the transition $(J,K)=(18,-7\rightarrow 19,-8)$ at 4.085 THz and $(J,K)=(15,-5\rightarrow 15,-4)$ at 4.125 THz with a variation $(1\rightarrow2)$ in the torsional state $\upsilon_T$. The transition frequencies were retrieved from the measurements of the pump beam frequencies provided by the wavelength meter (Bristol model 721), and were confirmed by a one-to-one comparison between the experimental data and the methanol spectrum, retrieved from the JPL molecular database (red line in Fig. 6a inset). Fig. 6a inset shows a 40 GHz mode-hop-free operation, that can be replicated along the whole generation bandwidth. All the line centers have been assigned and related to the amplitude of the DC signal used to tune the diode external cavity, as shown in fig. 6a, allowing a precise calibration of the frequency/voltage curve.     \\
Measurements performed with smaller frequency steps provide an overview on different line groups, with higher resolution. As an example, the five lines highlighted by the dashed rectangle in Fig. 6a inset are shown in Fig. 6b inset. In this 2 GHz wide scan, a linear calibration coefficient of 245 MHz/V can be retrieved, as shown in Fig. 6b.
\begin{figure}[tbph]
\begin{center}
\includegraphics[scale=0.18, keepaspectratio]{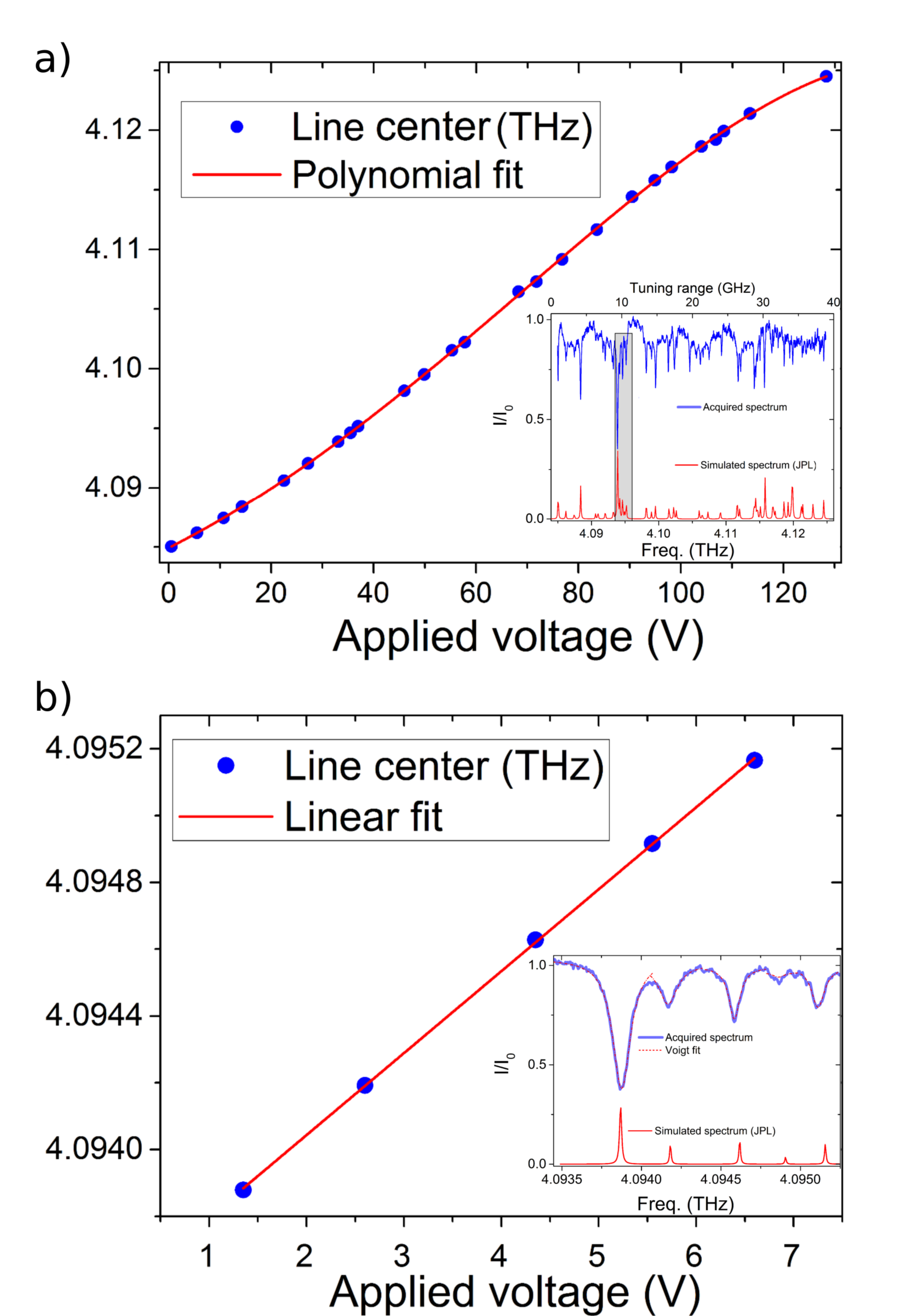}
\end{center}
\caption{a) center frequencies retrieved for the transitions shown in fig. 3a (replicated in the inset) b) Linear calibration of the piezo scan between 4.0935 and 4.0955 THz, a coefficient of 245 MHz/V is retrieved.} \label{fig:A}
\end{figure} 

\section{Spectral lines analysis}
The DC voltage signal used to achieve the high-resolution frequency scans was calibrated comparing the acquired methanol spectrum with the corresponding line-pattern provided by the JPL molecular database. The spectroscopic cell, providing a pressure stability with drifts reduced to less than 1\% in one hour, was sealed with a 2 mm thick TPX window and a 1 mm PP window. The PP filter was removed from the Golay-cell entrance window during the spectroscopic measurements. \\
When a laser beam of a given intensity $I_0$ passes through an absorbing material, the optical power density I, measured after a path $L$ within the sample, is reduced according to the Lambert-Beer law:\\
\begin{eqnarray}
I=I_0e^{-\alpha(\nu,p,T)L} ,
\end{eqnarray}
where $\alpha(\nu,p,T)$ represents the absorption coefficient and, for a gas sample, results from the convolution of a temperature-determined Gaussian profile (the Doppler contribution) with a Lorentzian term that takes into account the collisional broadening. At room temperature, lowering the vapor pressure to a few hundred Pa, the collisional width leaves off his dominant role, and a Voigt profile has to be considered to take into account both the Gaussian and the Lorentzian contributions.\\
The central frequencies of each spectral line are retrieved by fitting the experimental data to the following expression:
\begin{eqnarray}
-ln\left( \frac{I}{I_0} \right)= C+A \frac{2lnW_L}{\pi^{3/2}W_D} \cdot  \int^{+\infty}_{-\infty}\frac{e^{-t^2}dt}{\left( \sqrt{ln2} \frac{W_L}{W_D}\right)^2 + \left(\sqrt{4ln2} \frac{\nu-\nu_c}{W_D}-t^2\right)^2 }
\end{eqnarray}
Where $W_D$ and $W_L$ are the FWHM of the Doppler and Lorentzian contributions, respectively.

\end{document}